\documentclass[11pt,a4paper]{article}

\usepackage{epsfig,amsmath,amssymb,cite}

\begin{document}

\title{Thin--shell wormholes with a generalized Chaplygin gas } 
\author{Ernesto F. Eiroa$^{1, 2}$\thanks{e-mail: eiroa@iafe.uba.ar}\\
{\small $^1$ Instituto de Astronom\'{\i}a y F\'{\i}sica del Espacio, C.C. 67, 
Suc. 28, 1428, Buenos Aires, Argentina}\\
{\small $^2$ Departamento de F\'{\i}sica, Facultad de Ciencias Exactas y 
Naturales,} \\ 
{\small Universidad de Buenos Aires, Ciudad Universitaria Pab. I, 1428, 
Buenos Aires, Argentina}} 

\maketitle

\begin{abstract}
In this article, spherically symmetric thin--shell wormholes supported by a generalized Chaplygin gas are constructed and their stability under perturbations preserving the symmetry is studied. Wormholes with charge and with a cosmological constant are analyzed and the results are compared with those obtained for the original Chaplygin gas, which was considered in a previous work. For some values of the parameters, one stable configuration is also present and a new extra unstable solution is found.\\

\noindent 
PACS number(s): 04.20.Gz, 04.40.Nr, 98.80.Jk

\end{abstract}

\section{Introduction}\label{intro} 

Traversable Lorentzian wormholes \cite{motho} are solutions of the equations of gravitation representing geometries which have a throat that connects two regions of the same universe or two separate universes \cite{motho, visser}. For static wormholes, the throat is a minimal area surface satisfying a flare-out condition \cite{ hovis1}. In the context of the general relativity theory of gravitation, wormholes must be threaded by exotic matter that violates the null energy condition \cite{motho, visser, hovis1, hovis2}. The amount of exotic matter needed around the throat can be made arbitrarily small by an  appropriate choice of  the geometry of the wormhole \cite{viskardad}, although it may be at the expense of
large stresses at the throat \cite{dil,lst}.\\

Thin--shell wormholes are mathematically constructed by cutting and pasting two manifolds \cite{visser, mvis} to form a new one, which is geodesically complete and it has a shell placed  in the joining surface. The exotic matter required for their existence can be located at the shell, with normal matter outside this surface. Stability studies of thin--shell wormholes under perturbations preserving the original symmetries have been carried out in several works.  A linearized  stability analysis  of a thin--shell wormhole made by joining two Schwarzschild geometries was done in Ref. \cite{poisson}. The same method was applied to  wormholes constructed using branes with negative tensions \cite{barcelo}, and to transparent spherically symmetric thin--shells and wormholes \cite{ishak}. Later, it was extended to Reissner--Nordstr\"{o}m thin--shell wormholes \cite{eirom}, and to wormholes with a cosmological constant \cite{lobo}. Dynamical thin--shell wormholes were considered in Ref. \cite{lobo2}. The stability and energy conditions for five dimensional thin--shell wormholes with spherical symmetry in Einstein--Maxwell theory with a Gauss--Bonnet term were studied in Ref. \cite{marc}, while thin--shell wormholes in dilaton gravity were analyzed in Refs. \cite{dil, eir}. Thin--shell wormholes associated with cosmic strings have been treated in Refs. \cite{eisi}. Other recent related articles can be found in Refs. \cite{other}. \\

Observational data suggest an accelerated expansion of the Universe \cite{acc}, which within general relativity implies that the strong energy condition should be violated. Several models for the matter leading to this scenario have been proposed \cite{matt}. One of them is the Chaplygin gas \cite{chap}, a perfect fluid with an equation of state of the form $p\rho =-A$, where $A$ is a positive constant. The Chaplygin gas has the property that the squared sound velocity is always positive, even in the case of exotic matter. Although it was introduced for phenomenological reasons, not related with cosmology \cite{aero}, this equation of state can be obtained from string theory \cite{brane}. Models of exotic matter of interest in cosmology have already  been considered in wormhole construction. Wormholes supported by phantom energy, with equation of state $p=\omega \rho $, where $\omega<-1$, have been analyzed by several authors \cite{phantom}. A generalized Chaplygin gas, with an equation of state $p\rho^\alpha=-A$ which has two parameters $A>0$ and $0<\alpha\leq 1$, has been used in Ref. \cite{lobo73} as the exotic matter supporting a wormhole of the Morris--Thorne type \cite{motho}; there, the wormhole metric was matched to an exterior vacuum metric to keep the exotic matter within a finite region of space. Instead, in a thin--shell wormhole the exotic matter can be  restricted from the beginning to the shell located at the joining surface, as was previously done for the original Chaplygin gas ($\alpha =1$) in Ref. \cite{chaply}. Recently, other authors \cite{chapnew} have also considered wormholes with a Chaplygin gas. The purpose of the present work is the study of spherically symmetric thin--shell wormholes with matter in the form of a generalized Chaplygin gas. In Sec. 2, the equations that give the possible radii of wormholes  and determine their stability for a general class of geometries are obtained. In Secs. 3 and 4, the formalism is applied to Reissner--Nordstr\"om wormholes and to wormholes with a cosmological constant, respectively. Finally, in Sec. 5, the conclusions of this work are summarized. Units such that $c=G=1$ are adopted.\\

\section{General equations}\label{tswh}

We start from the spherically symmetric metric
\begin{equation} 
ds^2=-f(r)dt^2+f(r)^{-1}dr^2+h(r) (d\theta ^2+\sin^2\theta d\varphi^2), 
\label{e1}
\end{equation}
where $r>0$ is the radial coordinate, $0\le \theta \le \pi$ and $0\le \varphi<2\pi $ are the angular coordinates, $h(r)$ is always positive and $f(r)$ is a positive function from a given radius. We take a radius $a$, greater than the event horizon radius $r_{h}$ if the geometry (\ref{e1}) has any, to avoid the presence of horizons and singularities, we cut two identical copies of the region with $r\geq a$:
\begin{equation} 
\mathcal{M}^{\pm }=\{X^{\alpha }=(t,r,\theta,\varphi)/r\geq a\},  \label{e2}
\end{equation}
and we paste them at the hypersurface
\begin{equation} 
\Sigma \equiv \Sigma ^{\pm }=\{X/F(r)=r-a=0\},  \label{e3}
\end{equation}
to create a new manifold $\mathcal{M}=\mathcal{M}^{+}\cup \mathcal{M}^{-}$. If $h'(a)>0$ (condition of flare-out), this construction creates a geodesically complete manifold representing a wormhole with two regions connected by a throat of radius $a$, where the surface of minimal area is located. On this manifold it is possible to define a new radial coordinate $l=\pm \int_{a}^{r}\sqrt{1/f(r)}dr$ that represents the proper radial distance to the throat, which is situated at $l=0$; the plus and minus signs corresponding, respectively, to $\mathcal{M}^{+}$ and $\mathcal{M}^{-}$. On the wormhole throat $\Sigma $, which is a synchronous timelike hypersurface, we  define coordinates $\xi ^{i}=(\tau ,\theta,\varphi )$, with $\tau $ the proper time on the shell. We use the  Darmois-Israel formalism \cite{daris} and we let the throat radius be a function of time $a(\tau)$. The second fundamental forms (or extrinsic curvature) associated with the two sides of the shell are given by:
\begin{equation} 
K_{ij}^{\pm }=-n_{\gamma }^{\pm }\left. \left( \frac{\partial ^{2}X^{\gamma
} } {\partial \xi ^{i}\partial \xi ^{j}}+\Gamma _{\alpha \beta }^{\gamma }
\frac{ \partial X^{\alpha }}{\partial \xi ^{i}}\frac{\partial X^{\beta }}{
\partial \xi ^{j}}\right) \right| _{\Sigma },  
\label{e4}
\end{equation}
where $n_{\gamma }^{\pm }$ are the unit normals ($n^{\gamma }n_{\gamma }=1$) to $\Sigma $ in $\mathcal{M}$:
\begin{equation} 
n_{\gamma }^{\pm }=\pm \left| g^{\alpha \beta }\frac{\partial F}{\partial
X^{\alpha }}\frac{\partial F}{\partial X^{\beta }}\right| ^{-1/2}
\frac{\partial F}{\partial X^{\gamma }}.  
\label{e5}
\end{equation}
Adopting the orthonormal basis $\{ e_{\hat{\tau}}=e_{\tau }, e_{\hat{\theta}}=a^{-1}e_{\theta }, e_{\hat{\varphi}}=(a\sin \theta )^{-1} e_{\varphi }\} $, we obtain
\begin{equation} 
K_{\hat{\theta}\hat{\theta}}^{\pm }=K_{\hat{\varphi}\hat{\varphi}}^{\pm
}=\pm \frac{h'(a)}{2h(a)}\sqrt{f(a)+\dot{a}^2},
\label{e6}
\end{equation}
and
\begin{equation} 
K_{\hat{\tau}\hat{\tau}}^{\pm }=\mp \frac{f'(a)+2\ddot{a}}{2\sqrt{f(a)+\dot{a}^2}},
\label{e7}
\end{equation}
where a prime and the dot represent, respectively, the derivatives with respect to $r$ and $\tau$. With the definitions $[K_{_{\hat{\imath}\hat{\jmath}}}]\equiv K_{_{\hat{\imath}\hat{\jmath}}}^{+}-K_{_{\hat{\imath}\hat{\jmath}}}^{-}$, $K=tr[K_{\hat{\imath}\hat{\jmath }}]=[K_{\; \hat{\imath}}^{\hat{\imath}}]$ and introducing  the surface stress-energy tensor $S_{_{\hat{\imath}\hat{\jmath} }}={\rm diag}(\sigma ,p_{\hat{\theta}},p_{\hat{\varphi}})$, where $\sigma$ is the surface energy density and $p_{\hat{\theta}}$, $p_{\hat{\varphi}}$ are the transverse pressures, we have from the Einstein equations on the shell (or Lanczos equations):
\begin{equation} 
-[K_{\hat{\imath}\hat{\jmath}}]+Kg_{\hat{\imath}\hat{\jmath}}=8\pi 
S_{\hat{\imath}\hat{\jmath}},
\label{e8}
\end{equation}
that 
\begin{equation} 
\sigma=-\frac{\sqrt{f(a)+\dot{a}^2}}{4\pi }\frac{h'(a)}{h(a)},
\label{e9}
\end{equation}
and
\begin{equation}
p=p_{\hat{\theta}}=p_{\hat{\varphi}}=\frac{\sqrt{f(a)+\dot{a}^2}}{8\pi}\left[ \frac{2\ddot{a}+f'(a)}{f(a)+\dot{a}^2}+\frac{h'(a)}{h(a)}\right] .
\label{e10}
\end{equation}
The negative sign in Eq. (\ref{e9}) plus the flare-out condition $h'(a)>0$ implies that $\sigma <0$, indicating that the matter at the throat is exotic. We adopt a generalized Chaplygin gas as the exotic matter in the shell $\Sigma $. For this gas, the pressure has opposite sign to the energy density, resulting in a positive pressure. Then, the equation of state for the exotic matter at the throat can be written in the form
\begin{equation}
p=\frac{A}{|\sigma |^{\alpha}},
\label{e11} 
\end{equation} 
where $A>0$ and $0<\alpha\le 1$ are constants. When $\alpha=1$ the Chaplygin gas equation of state $p=-A/\sigma$ is recovered. Replacing Eqs. (\ref{e9}) and (\ref{e10}) in Eq. (\ref{e11}), we obtain the differential equation that should be satisfied by the throat radius of thin--shell wormholes threaded by exotic matter with the equation of state of a generalized Chaplygin gas:
\begin{equation}
\{ [2\ddot{a}+f'(a)]h(a)+[f(a)+\dot{a}^2]h'(a) \}[h'(a)]^{\alpha}-2A[4\pi h(a)]^{\alpha +1}[f(a)+\dot{a}^2]^{(1-\alpha)/2}=0.
\label{e12} 
\end{equation}
In particular, for static wormholes, the surface energy density and pressure are given  by
\begin{equation}
\sigma_{0}=-\frac{\sqrt{f(a_{0})}}{4\pi}\frac{h'(a_{0})}{h(a_{0})},
\label{e13}
\end{equation}
and
\begin{equation}
p_{0}=\frac{\sqrt{f(a_{0})}}{8\pi}\left[ \frac{f'(a_{0})}{f(a_{0})}+\frac{h'(a_{0})}{h(a_{0})}\right] .
\label{e14}
\end{equation}
From Eq. (\ref{e12}), the static solutions, if they exist, have a throat radius $a_{0}$ that should fulfill the equation
\begin{equation}
[f'(a_{0})h(a_{0})+f(a_{0})h'(a_{0})][h'(a_{0})]^{\alpha}-2A[4\pi h(a_{0})]^{\alpha +1}[f(a_{0})]^{(1-\alpha)/2}=0,
\label{e15} 
\end{equation} 
with the condition $a_{0}>r_{h}$ if the original metric has an event horizon. The existence of static solutions depends on the explicit form of the function $f$. The method developed in Ref. \cite{chaply} to study the stability of the static solutions under perturbations preserving the symmetry in the case of the Chaplygin gas is rather cumbersome to be extended to the generalized Chaplygin gas. Here we follow the standard ``potential" approach (both methods were shown equivalent for the metric given by Eq. (\ref{e1}) in the case of a linearized equation of state \cite{eir}). From Eqs. (\ref{e9}) and (\ref{e10}), it is easy to check the energy conservation equation:
\begin{equation}
\frac{d}{d\tau }\left( \sigma \mathcal{A}\right) +p\frac{d\mathcal{A}}{d\tau }=
\left\lbrace \left[ h'(a)\right]^2 -2h(a)h''(a)\right\rbrace \frac{\dot{a}\sqrt{f(a)+ \dot{a}^2}}{2h(a)},
\label{p1}
\end{equation}
where $\mathcal{A}=4\pi h(a)$ is the area of the wormhole throat. In Eq. (\ref{p1}), the first term in the left hand side represents the internal energy change of the throat and the second the work done by the internal forces of the throat, while the right hand side represents a flux. If $[h'(a)]^2 -2h(a)h''(a)=0$, the flux term is zero and Eq. (\ref{p1}) takes the form of a simple conservation equation. This occurs when $h(a)=C(a+D)^2$  ($C>0$ and $D$ constants) or  $h(a)=C$  (this case is unphysical, since there is no throat) \cite{eir}. It is straightforward to see that Eq. (\ref{p1}) can be written in the form
\begin{equation}
h(a)\dot{\sigma}+h'(a)\dot{a}(\sigma +p)=-\left\lbrace \left[ h'(a)\right]^2 -2h(a)h''(a)\right\rbrace
\frac{\dot{a}\sigma }{2 h'(a)},
\label{p2}
\end{equation}
which using that $\sigma '=\dot{\sigma }/\dot{a}$ gives
\begin{equation}
h(a)\sigma '+h'(a)(\sigma +p)+\left\lbrace \left[ h'(a)\right]^2 -2h(a)h''(a)\right\rbrace
\frac{\sigma }{2 h'(a)}=0.
\label{p3}
\end{equation}
The pressure $p$ is a function of $\sigma $ given by the equation of state, thus Eq. (\ref{p3}) is a first order differential that can be recast in the form $\sigma '(a)=\mathcal{F}(a, \sigma (a))$, for which a unique solution with a given initial condition always exists, provided that $\mathcal{F}$ has continuous partial derivatives. Then, Eq. (\ref{p3}) can be formally integrated to obtain $\sigma (a)$, so replacing $\sigma (a)$ in Eq. (\ref{e9}) and regrouping terms, the dynamics of the wormhole throat is completely determined by a single equation:
\begin{equation}
\dot{a}^{2}=-V(a),
\label{p4}
\end{equation}
with 
\begin{equation}
V(a)=f(a)-16\pi ^{2}\left[ \frac{h(a)}{h'(a)}\sigma (a)\right] ^{2}.
\label{p5}
\end{equation}
A Taylor expansion to second order of the potential $V(a)$ around the static solution yields:
\begin{equation}
V(a)=V(a_{0})+V'(a_{0})(a-a_{0})+\frac{V''(a_{0})}{2}
(a-a_{0})^{2}+O(a-a_{0})^{3}.
\label{p6}
\end{equation}
From Eq. (\ref{p5}) the first derivative of $V(a)$ is
\begin{equation}
V'(a)=f'(a)-32\pi ^{2}\sigma (a) \frac{h(a)}{h'(a)} \left\lbrace  \left[ 1-\frac{h(a)h''(a)}{[h'(a)]^2}\right] \sigma (a) +\frac{h(a)}{h'(a)}\sigma '(a)\right\rbrace ,
\label{p7}
\end{equation}
which using Eq. (\ref{p3}) takes the form
\begin{equation}
V'(a)=f'(a)+16\pi ^{2}\sigma (a)\frac{h(a)}{h'(a)} \left[ \sigma (a)+ 2 p(a) \right] .
\label{p8}
\end{equation}
The second derivative of the potential is
\begin{eqnarray}
V''(a)&=&f''(a)+16\pi ^{2}\left\lbrace \left[ \frac{h(a)}{h'(a)}\sigma '(a)+\left( 1-\frac{h(a)h''(a)}{[h'(a)]^2} \right)  \sigma (a)\right] \left[ \sigma (a)+2p(a)\right] \right. \nonumber
\\ 
& & \left. +\frac{h(a)}{h'(a)}\sigma (a) \left[ \sigma '(a)+2p'(a)\right] \right\rbrace  .
\label{p9}
\end{eqnarray}
From Eq. (\ref{e11}) we have that $p'(a)=A\alpha|\sigma (a)|^{-\alpha -1}\sigma '(a)=\alpha p(a)\sigma '(a)/|\sigma (a)|$, then $\sigma '(a)+2p'(a)=\sigma '(a)[1+2\alpha p(a)/|\sigma (a)|]$, and using Eq. (\ref{p3}) again, we obtain 
\begin{equation}
V''(a)=f''(a)-8\pi ^{2}\left\lbrace \left[ \sigma (a)+2p(a)\right] ^{2} +2\sigma (a)\left[ \left(  \frac{3}{2}-\frac{h(a)h''(a)}{[h'(a)]^2}\right) \sigma (a)+p(a)\right] \left[ 1+2\frac{\alpha p(a)}{|\sigma (a)|}\right] \right\rbrace .
\label{p10}
\end{equation}
Using Eqs. (\ref{e13}) and (\ref{e14}), it is not difficult to see that $V(a_{0})=V'(a_{0})=0$, so the potential is
\begin{equation}
V(a)=\frac{1}{2}V''(a_{0})(a-a_{0})^{2}+O[(a-a_{0})^{3}],
\label{p11}
\end{equation}
with 
\begin{eqnarray}
V''(a_{0})&=& f''(a_{0})+\frac{(\alpha -1)[f'(a_{0})]^2}{2f(a_{0})}+\left[ \frac{(1-\alpha )h'(a_{0})}{2h(a_{0})}+\frac{\alpha h''(a_{0})}{h'(a_{0})}\right] f'(a_{0}) \nonumber \\
& & +(\alpha +1)\left[ \frac{h''(a_{0})}{h(a_{0})} -\left( \frac{h'(a_{0})}{h(a_{0})} \right) ^{2}\right] f(a_{0}).
\label{p12}
\end{eqnarray}
The wormhole is stable under radial perturbations if and only if $V''(a_{0})>0$.

\section{Reissner-Nordstr\"om wormholes}\label{rn}

The Reissner--Norsdtr\"om geometry, which represents a spherically symmetric charged object, have metric functions
\begin{equation}
f(r)=1-\frac{2M}{r}+\frac{Q^2}{r^2},\hspace{1cm} h(r)=r^2,
\label{rn1} 
\end{equation} 
where $M$ is the mass and $Q$ is the charge. For $0<|Q|<M$ this geometry has two horizons with radii
\begin{equation}
r_{\pm}=M\pm \sqrt{M^{2}-Q^{2}},
\label{rn2} 
\end{equation} 
where the minus sign corresponds to the inner one and the plus sign to the outer event horizon. When $|Q|=M$ the two horizons merge into one, and if $|Q|>M$ no horizons are present and there is a naked singularity in $r_{s}=0$. If $Q=0$ the Schwarzschild geometry is obtained, which has a horizon with radius $r_{h}=2M$. The thin--shell wormholes constructed from the Reissner--Norsdtr\"om metric consist of a charged surface of exotic matter (the throat) with an electric field in vacuum outside it. When $|Q|\le M$ the throat radius $a_{0}$ should be taken greater than $r_{h}=r_{+}$ so that the manifold  $\mathcal{M}$ has no horizons. If $|Q|>M$ the condition $a_{0}>0$ removes the naked singularity. By replacing Eq. (\ref{rn1}) in Eqs. (\ref{e13}) and (\ref{e14}), the energy density and pressure at the throat are obtained:
\begin{equation}
\sigma _{0}=-\frac{\sqrt{a_{0}^{2}-2Ma_{0}+{Q^2}}}{2\pi a_{0}^{2}},
\label{rn3} 
\end{equation}
and
\begin{equation}
p_{0}=\frac{a_{0}-M}{4\pi a_{0}\sqrt{a_{0}^{2}-2Ma_{0}+{Q^2}}}.
\label{rn4} 
\end{equation}
\begin{figure}[t!]
\begin{center}
\vspace{0cm}
\includegraphics[width=12cm]{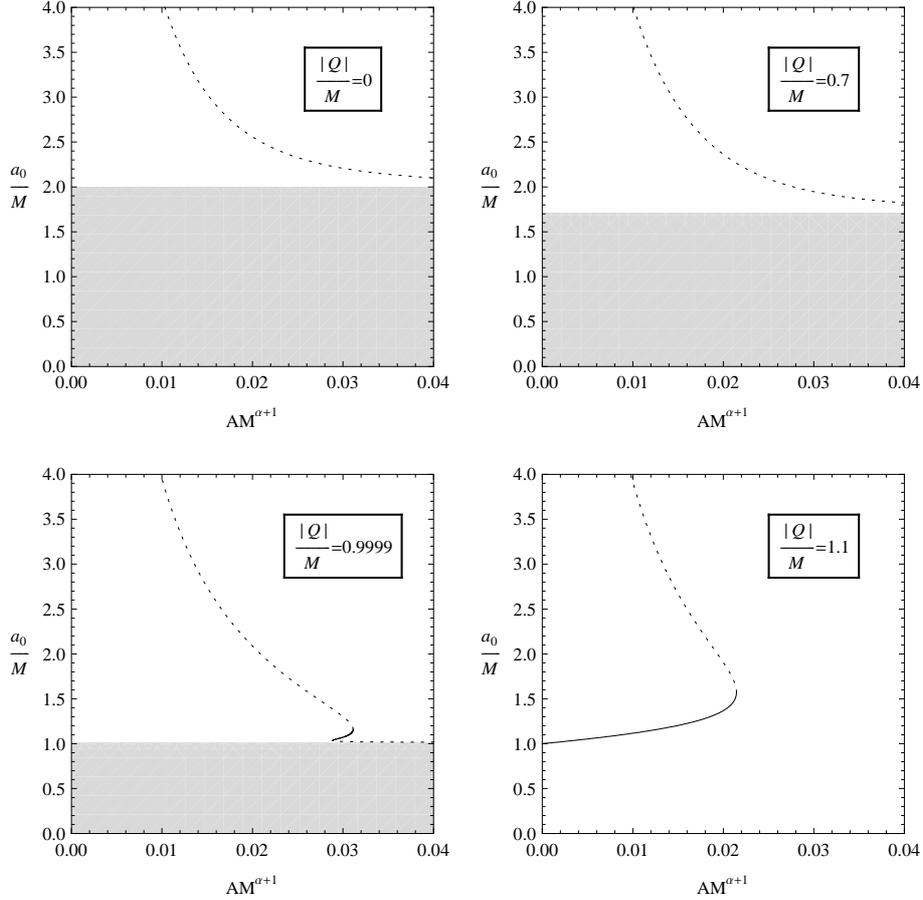}
\vspace{-0.5cm}
\end{center} 
\caption{Reissner--Nordstr\"{o}m wormholes supported by a generalized Chaplygin gas with $\alpha =0.2$: the solid curves represent the static solutions with throat radius $a_{0}$ which are stable under radial perturbations for given parameters $A$, $M$ and $Q$, and the dotted curves those unstable under radial perturbations. The gray zones are unphysical, corresponding to a throat radius smaller than the horizon radius of the original manifold.  }
\label{frn1}
\end{figure} 
\begin{figure}[t!]
\begin{center}
\vspace{0cm}
\includegraphics[width=12cm]{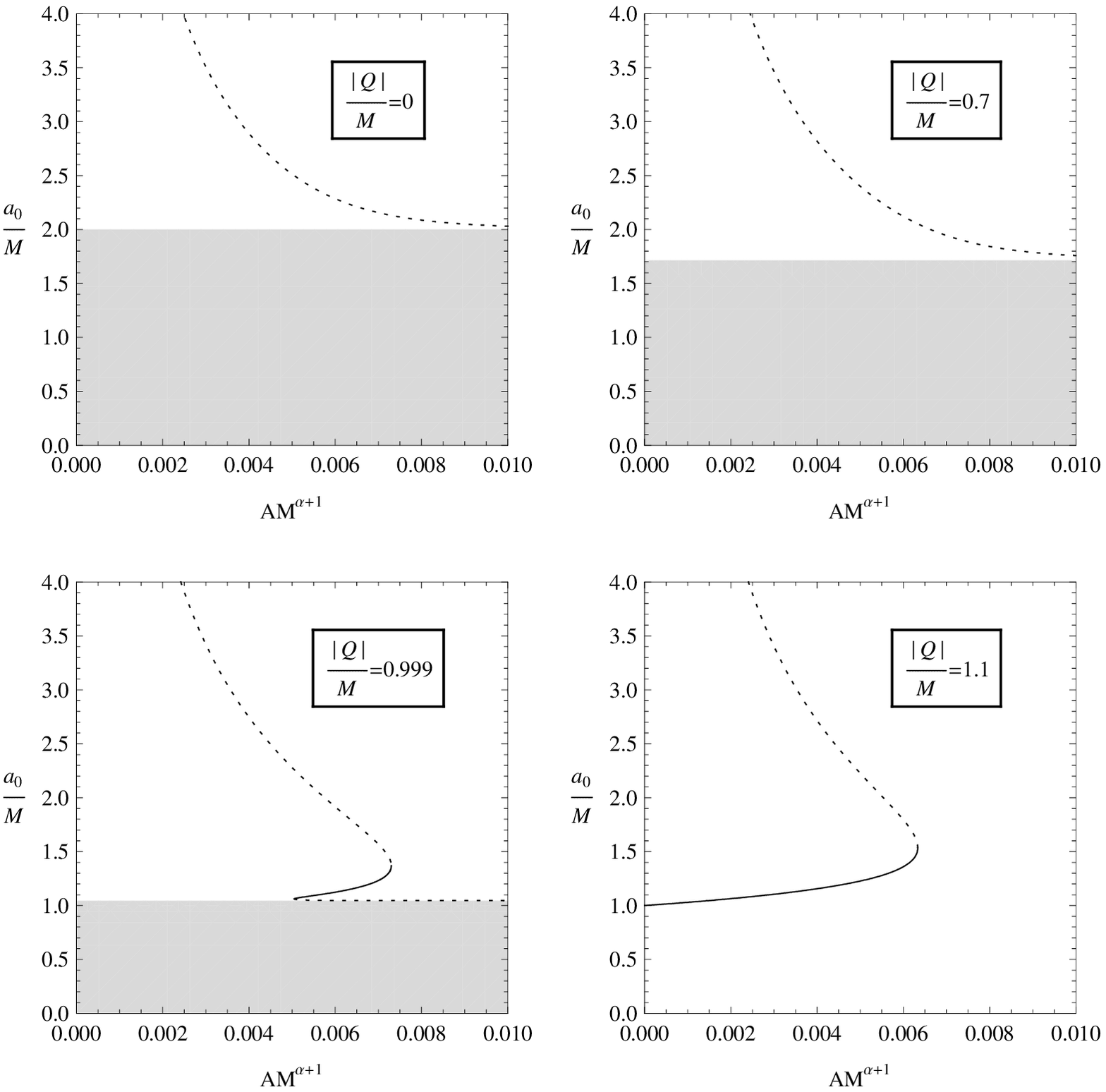}
\vspace{-0.5cm}
\end{center} 
\caption{Reissner--Nordstr\"{o}m wormholes supported by a generalized Chaplygin gas with $\alpha =0.6$: the solid curves represent the static solutions with throat radius $a_{0}$ which are stable under radial perturbations for given parameters $A$, $M$ and $Q$, and the dotted curves those unstable under radial perturbations. The gray zones are unphysical, corresponding to a throat radius smaller than the horizon radius of the original manifold. }
\label{frn2}
\end{figure}
\begin{figure}[t!]
\begin{center}
\vspace{0cm}
\includegraphics[width=12cm]{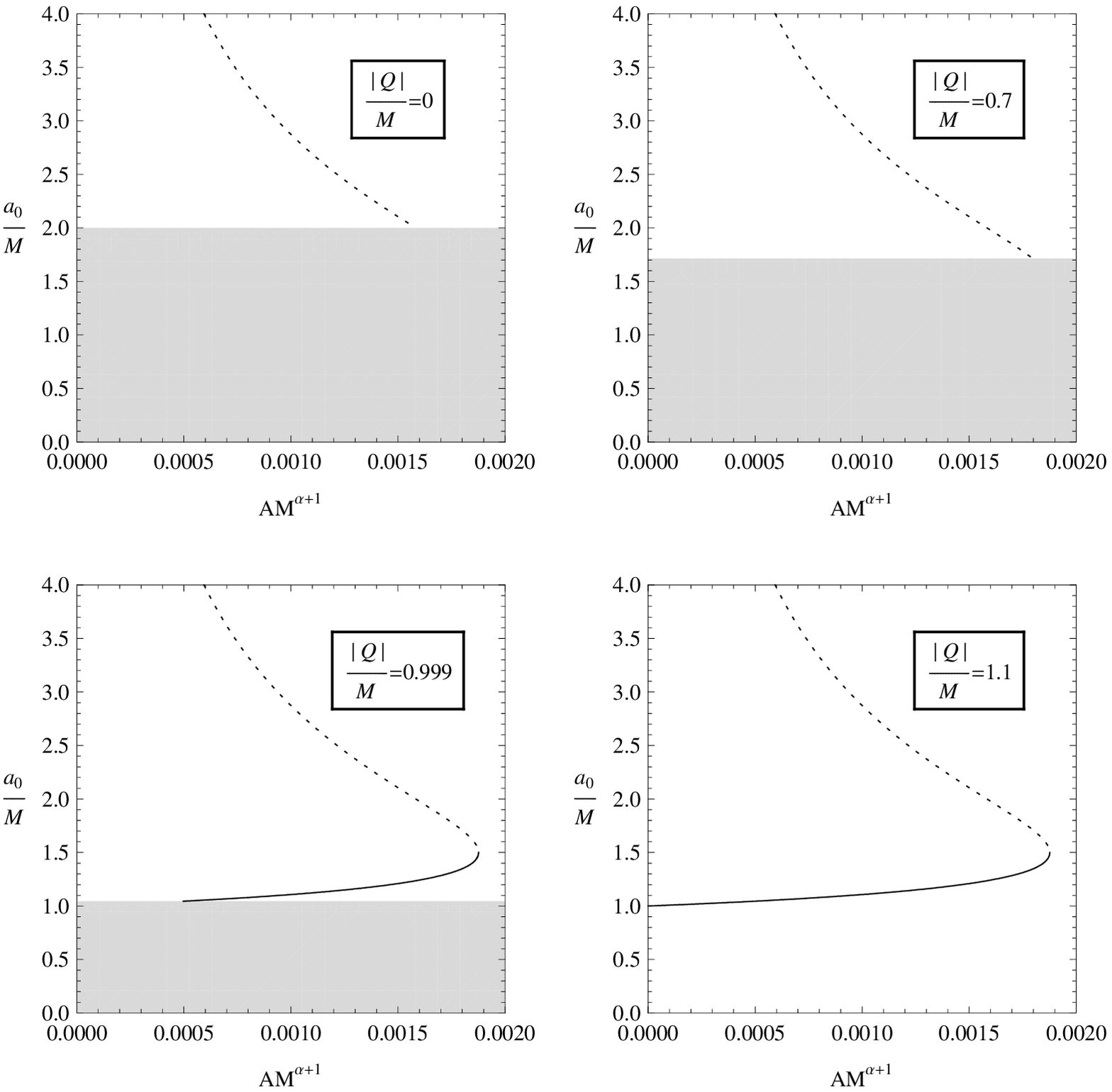}
\vspace{-0.5cm}
\end{center} 
\caption{Reissner--Nordstr\"{o}m wormholes supported by a Chaplygin gas ($\alpha =1)$: the solid curves represent the static solutions with throat radius $a_{0}$ which are stable under radial perturbations for given parameters $A$, $M$ and $Q$, and the dotted curves those unstable under radial perturbations. The gray zones are unphysical, corresponding to a throat radius smaller than the horizon radius of the original manifold. }
\label{frn3}
\end{figure}
The throat radius should satisfy the equation
\begin{equation}
a_{0}-M-2A(2\pi )^{\alpha +1}a_{0}^{2\alpha +1}(a_{0}^{2}-2Ma_{0}+Q^{2})^{(1-\alpha)/2}=0.
\label{rn5}
\end{equation} 
If $\alpha =1$ the equation is cubic and it can be solved analytically (see Ref. \cite{chaply}) to obtain $a_{0}$, while for other values of $\alpha $ it should be solved numerically. For the stability analysis of the solutions, we obtain from Eqs. (\ref{p12}) and (\ref{rn1}) the second derivative of the potential, which is 
\begin{equation}
V''(a_{0})=\frac{2 \left\{ -(1 + \alpha ) a_{0}^{3} + M (3 + 4 \alpha ) a_{0}^{2} - \left[ 2 Q^2 \alpha + 3 M^2 (1 + \alpha )\right] a_{0} + M Q^2 (1 + 2 \alpha ) \right\} }
{a_{0}^{3}(a_{0}^{2} - 2 M a_{0} + Q^{2})}.
\label{rn6}
\end{equation} 
The solutions of Eq. (\ref{rn5}) correspond to stable wormholes if replaced in Eq. (33) they satisfy $V''(a_{0})>0$. The calculations were done with standard software and the results for representative values of the parameters are shown in Figs. \ref{frn1}, \ref{frn2} and \ref{frn3}. When $0<\alpha <1$ we can summarize the results of Figs. \ref{frn1} and \ref{frn2} as follows: 
\begin{enumerate}
\item When $0\le |Q|<1$ and $|Q|$ is not very close to $1$, for any value of $AM^{\alpha +1}$ there is always one unstable solution. The throat radius $a_{0}/M$ decreases with $AM^{\alpha +1}$ and tends to the horizon radius $r_{h}/M$ of the original manifold for large values of $AM^{\alpha +1}$.
\item If $|Q|<1$ and $|Q|$ is very close to $1$, for small values of $AM^{\alpha +1}$ there is one unstable solution. For intermediate values of $AM^{\alpha +1}$ there are three solutions: the larger and the smaller ones are unstable and the middle one is stable. For large values of $AM^{\alpha +1}$ there is one unstable solution, which is very close to the horizon of the original manifold.
\item When $|Q|>1$, for small values of $AM^{\alpha +1}$ there are two solutions: the larger one is unstable and the smaller one is stable, while for large values of $AM^{\alpha +1}$ there are no solutions.
\end{enumerate}
The case $\alpha =1$ corresponds to the Chaplygin gas studied in detail in a previous work \cite{chaply}. For comparison, the results are shown in Fig. \ref{frn3}. We see that
\begin{enumerate}
\item When $0\le |Q|<1$ and $|Q|$ is not very close to $1$, for small values of $AM^{\alpha +1}$ there is always one unstable solution. The throat radius $a_{0}/M$ decreases with $AM^{\alpha +1}$ and it cuts to the horizon radius $r_{h}/M$ of the original manifold for a finite value of $AM^{\alpha +1}$, so there are no solutions for large values of $AM^{\alpha +1}$.
\item If $|Q|<1$ and $|Q|$ is very close to $1$, for small values of $AM^{\alpha +1}$ there is one unstable solution. For intermediate values of $AM^{\alpha +1}$ there are two solutions: the larger one is unstable and the smaller one is stable. For large values of $AM^{\alpha +1}$ there are no solutions.
\item When $|Q|>1$, for small values of $AM^{\alpha +1}$ there are two solutions: the larger one is unstable and the smaller one is stable, while for large values of $AM^{\alpha +1}$ there are no solutions.
\end{enumerate}
We see that the main difference between the cases $0<\alpha <1$ and $\alpha =1$ is that in the former case there is an extra unstable solution for some values of the parameters.

\section{Wormholes with a cosmological constant}\label{schw-ds}

The vacuum solution of the Einstein equations with a cosmological constant has metric functions
\begin{equation}
f(r)=1-\frac{2M}{r}-\frac{\Lambda}{3}r^2, \hspace{1cm} h(r)=r^2,
\label{cc1} 
\end{equation} 
where $M$ is the mass and $\Lambda $ is the cosmological constant. If $\Lambda M^{2}>1/9$ the function $f(r)$ is always negative, so we take $\Lambda M^{2}\le 1/9$. When $0<\Lambda M^{2}\le 1/9$, i.e. the Schwarzschild--de Sitter case, the geometry has two horizons, which are placed at
\begin{equation}
r_{h}^{dS}=\frac{-1+i\sqrt{3}-(1+i\sqrt{3})\left(-3\sqrt{\Lambda}M+i\sqrt{1-9\Lambda M^{2}}
\right)^{2/3}}{2\sqrt{\Lambda}\left(-3\sqrt{\Lambda}M+i\sqrt{1-9\Lambda M^{2}}\right)^{1/3}},
\label{ds2} 
\end{equation} 
\begin{equation}
r_{c}^{dS}=\frac{1+\left(-3\sqrt{\Lambda}M+i\sqrt{1-9\Lambda M^{2}}\right)^{2/3}}
{\sqrt{\Lambda}\left(-3\sqrt{\Lambda}M+i\sqrt{1-9\Lambda M^{2}}\right)^{1/3}}.
\label{ds3} 
\end{equation}
The event horizon radius $r_{h}^{dS}$ is a continuous and increasing function of $\Lambda $, with  $r_{h}^{dS}\rightarrow 2M$ when $\Lambda \rightarrow 0^{+}$ and $r_{h}^{dS}=3M$ when $\Lambda M^{2}=1/9$. The cosmological horizon radius $r_{c}^{dS}$ is a continuous and decreasing function of $\Lambda $, with $r_{c}^{dS}\rightarrow +\infty$ when $\Lambda \rightarrow 0^{+}$ and $r_{c}^{dS}=3M$ when $\Lambda M^{2}=1/9$. If $\Lambda =0$, the Schwarzschild geometry with horizon radius $r_{h}^{S}=2M$ is obtained. When $\Lambda <0$, i.e. the Schwarzschild--anti de Sitter case, the event horizon is placed at 
\begin{equation}
r_{h}^{AdS}=\frac{1-\left(-3\sqrt{|\Lambda|}M+\sqrt{1+9|\Lambda|M^{2}}\right)^{2/3}}
{\sqrt{|\Lambda|}\left(-3\sqrt{|\Lambda|}M+\sqrt{1+9|\Lambda|M^{2}}\right)^{1/3}}.
\label{ads1} 
\end{equation}
The horizon radius $r_{h}^{AdS}$ is a continuous and increasing function of $\Lambda$, with values in the interval $0<r_{h}<2M$, with $r_{h}^{AdS}=0$ when $\Lambda\rightarrow -\infty$ and $r_{h}^{AdS}\rightarrow 2M$ when $\Lambda \rightarrow 0^{-}$. Thin--shell wormholes constructed from Eq. (\ref{cc1}) have a throat of exotic matter and vacuum with a cosmological constant outside it. If $0<\Lambda M^{2}<1/9$ the wormhole throat radius should be taken in the range $r_{h}^{dS}<a_{0}<r_{c}^{dS}$, and if $\Lambda M^{2}=1/9$ the construction of the wormhole is not possible, because $r_{h}^{dS}=r_{c}^{dS}=3M$. If $\Lambda <0$ the wormhole throat radius $a_{0}$ should be greater than $r_{h}^{AdS}$. Using Eqs. (\ref{e13}) and (\ref{e14}), we obtain that the energy density and the pressure at the throat are given by
\begin{equation}
\sigma _{0}=-\frac{\sqrt{-\Lambda a_{0}^{3}+3a_{0}-6M}}{2\pi a_{0} \sqrt{3 a_{0}}},
\label{cc2} 
\end{equation}
and
\begin{figure}[t!]
\begin{center}
\vspace{0cm}
\includegraphics[width=12cm]{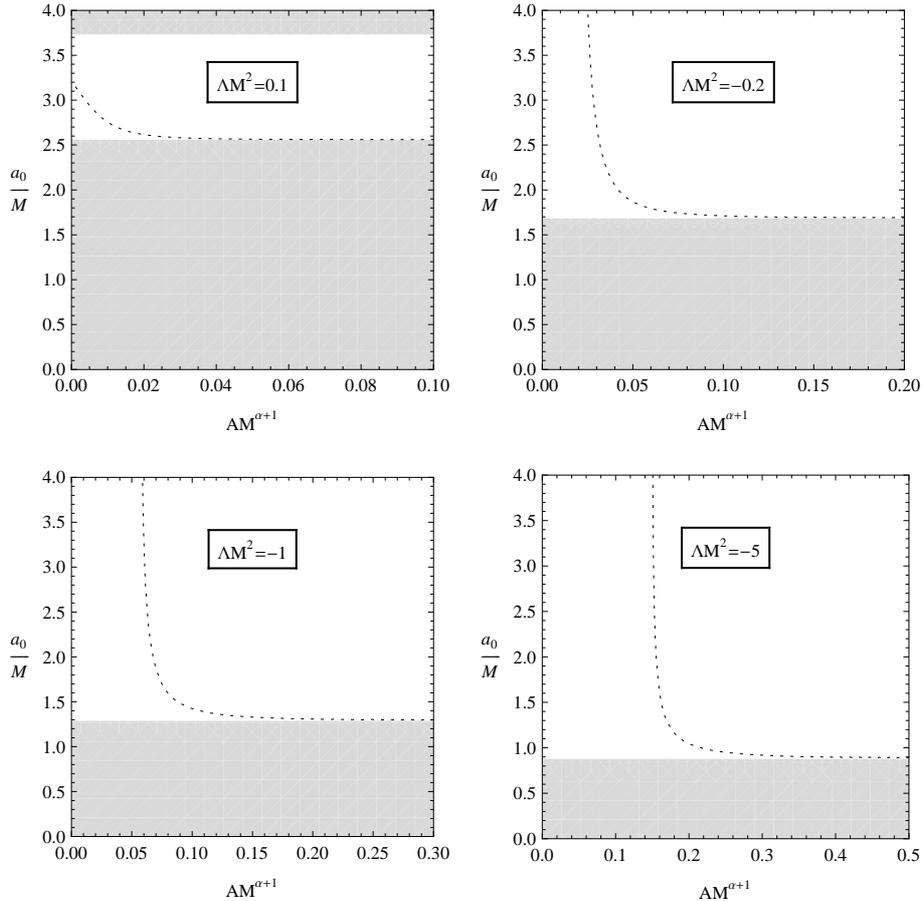}
\vspace{-0.5cm}
\end{center} 
\caption{Wormholes with a cosmological constant, supported by a generalized Chaplygin gas with $\alpha =0.2$: the solid curves represent the static solutions with throat radius $a_{0}$ which are stable under radial perturbations for given parameters $A$, $M$ and $\Lambda $, and the dotted curves those unstable under radial perturbations. The gray zones are unphysical, corresponding to a throat radius smaller than the horizon radius or (if $\Lambda >0$) larger than the cosmological horizon radius of the original manifold.  }
\label{fcc1}
\end{figure} 
\begin{figure}[t!]
\begin{center}
\vspace{0cm}
\includegraphics[width=12cm]{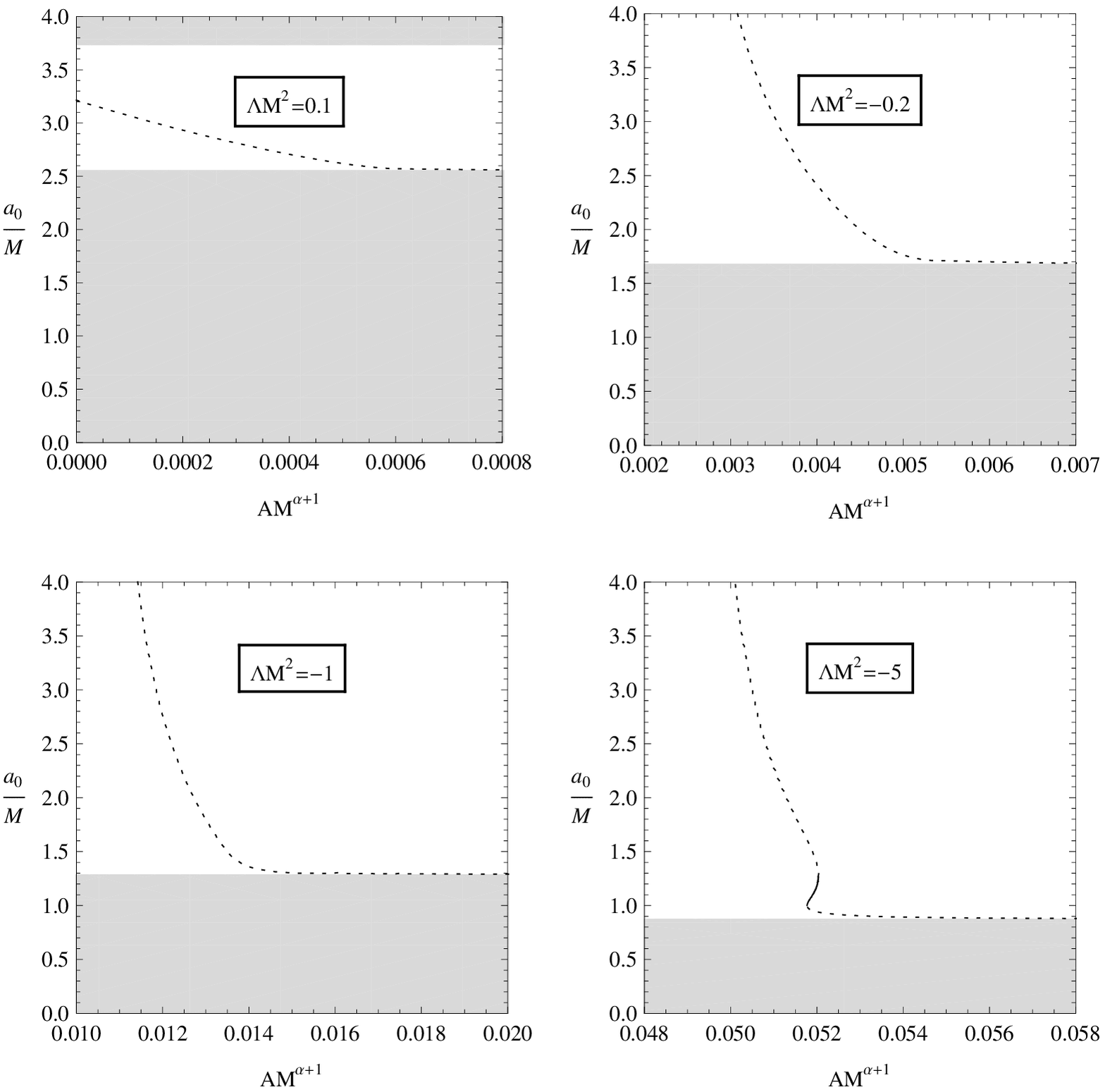}
\vspace{-0.5cm}
\end{center} 
\caption{Wormholes with a cosmological constant, supported by a generalized Chaplygin gas with $\alpha =0.9$: the solid curves represent the static solutions with throat radius $a_{0}$ which are stable under radial perturbations for given parameters $A$, $M$ and $\Lambda $, and the dotted curves those unstable under radial perturbations. The gray zones are unphysical, corresponding to a throat radius smaller than the horizon radius or (if $\Lambda >0$) larger than the cosmological horizon radius of the original manifold. }
\label{fcc2}
\end{figure}
\begin{figure}[t!]
\begin{center}
\vspace{0cm}
\includegraphics[width=12cm]{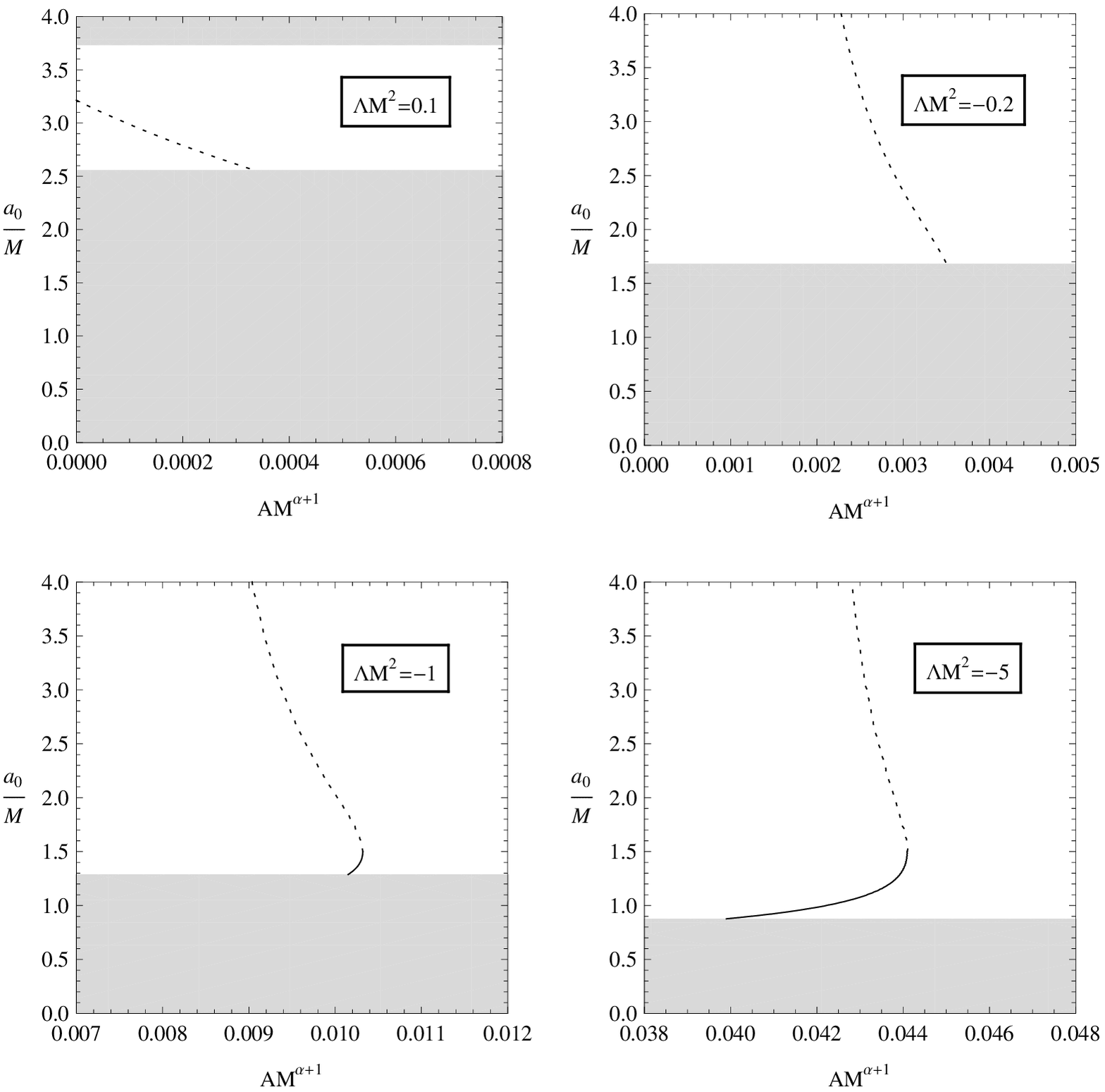}
\vspace{-0.5cm}
\end{center} 
\caption{Wormholes with a cosmological constant, supported by a Chaplygin gas ($\alpha =1)$: the solid curves represent the static solutions with throat radius $a_{0}$ which are stable under radial perturbations for given parameters $A$, $M$ and $\Lambda $, and the dotted curves those unstable under radial perturbations. The gray zones are unphysical, corresponding to a throat radius smaller than the horizon radius or (if $\Lambda >0$) larger than the cosmological horizon radius of the original manifold. }
\label{fcc3}
\end{figure}
\begin{equation}
p_{0}=\frac{-2 \Lambda a_{0}^{3}+3a_{0}-3M}{4 \pi a_{0} \sqrt{3a_{0}(-\Lambda a_{0}^{3}+3a_{0}-6M)}}.
\label{cc3} 
\end{equation} 
From Eq. (\ref{e15}), the throat radius $a_{0}$ should satisfy the equation
\begin{equation}
-\frac{2\Lambda}{3}a_{0}^{3}+a_{0}-M-2A(2\pi )^{\alpha+1}a_{0}^{3(\alpha +1)/2}\left( -\frac{\Lambda}{3}a_{0}^{3}+a_{0}-2M\right)^{(1-\alpha)/2}=0.
\label{cc4} 
\end{equation}
When $\alpha =1$ this equation is cubic in $a_{0}$ and it can be solved analytically \cite{chaply}, while for other values of $\alpha $ it should be solved numerically. By replacing Eq. (\ref{cc1}) in Eq. (\ref{p12}), we obtain 
\begin{equation}
V''(a_{0})=\frac{2 \left[ -2 \alpha \Lambda a_{0}^{4} + 3 M (-1 + 2 \alpha) \Lambda a_{0}^{3} + 3 (1 + \alpha) a_{0}^2 -  3 M (3 + 4 \alpha )a_{0} + 9 M^2 (1 + \alpha) \right] }{a_{0}^3 (\Lambda a_{0}^3 -3 a_{0} + 6 M )}.
\label{cc5}
\end{equation} 
The solutions of Eq. (\ref{cc4}) correspond to stable wormholes if $V''(a_{0})>0$.  The calculations were done with standard software and the results for representative values of the parameters are shown in Figs. \ref{fcc1}, \ref{fcc2} and \ref{fcc3}. When $0<\alpha <1$ we can summarize the results of Figs. \ref{fcc1} and \ref{fcc2} as follows: 
\begin{enumerate}
\item If $\Lambda > 0$, for any value of $AM^{\alpha +1}$ there is always one unstable solution. The throat radius $a_{0}/M$ decreases with $AM^{\alpha +1}$ and tends to the horizon radius $r_{h}^{dS}/M$ of the original manifold for large values of $AM^{\alpha +1}$. Except for the presence of the cosmological horizon, the results are very similar to the ones obtained for the Schwarzschild wormholes ($\Lambda =0$), which were studied in the previous section (uncharged case).
\item When $\Lambda <0$ and $|\Lambda|M^{2}$ is not large, for any value of $AM^{\alpha +1}$ there is one unstable solution. The throat radius $a_{0}/M$ decreases with $AM^{\alpha +1}$ and tends to the horizon radius $r_{h}^{AdS}/M$ of the original manifold for large values of $AM^{\alpha +1}$.
\item If $\Lambda <0$ and $|\Lambda|M^{2}$ is large, for small values of $AM^{\alpha +1}$ there is one unstable solution. For intermediate values of $AM^{\alpha +1}$ there are three solutions: the larger and the smaller ones are unstable and the middle one stable. For large values of $AM^{\alpha +1}$ there is one unstable solution, which is very close to the horizon $r_{h}^{AdS}/M$ of the original manifold.
\end{enumerate}
The case $\alpha =1$ corresponds to the Chaplygin gas wormholes previously studied in Ref. \cite{chaply}. The results are shown in Fig. \ref{fcc3} for comparison. We have that
\begin{enumerate}
\item When $\Lambda > 0$, for small values of $AM^{\alpha +1}$ there is always one unstable solution. The throat radius $a_{0}/M$ decreases with $AM^{\alpha +1}$ and it cuts to the horizon radius $r_{h}^{dS}/M$ of the original manifold for a finite value of $AM^{\alpha +1}$, so there are no solutions for large values of $AM^{\alpha +1}$. Again, except for the presence of the cosmological horizon, the results are very similar to the ones obtained for the Schwarzschild wormholes ($\Lambda =0$), which were analyzed in the previous section (uncharged case).
\item If $\Lambda < 0$ and $|\Lambda| M^{2}$ is not large, for small values of $AM^{\alpha +1}$ there is always one unstable solution. The throat radius $a_{0}/M$ decreases with $AM^{\alpha +1}$ and it cuts to the horizon radius $r_{h}^{AdS}/M$ of the original manifold for a finite value of $AM^{\alpha +1}$, so there are no solutions for large values of $AM^{\alpha +1}$.
\item  When $\Lambda <0$ and $|\Lambda| M^{2}$ is large, for small values of $AM^{\alpha +1}$ there is one unstable solution. For intermediate values of $AM^{\alpha +1}$ there are two solutions: the larger one is unstable and the smaller one is stable. For large values of $AM^{\alpha +1}$ there are no solutions.
\end{enumerate}
We can see that the main difference between the cases $0<\alpha <1$ and $\alpha =1$ is that in the former case there are values of the parameters for which exist one unstable additional solution for anti--de Sitter wormholes.

\section{Conclusions}\label{conclu}

In this paper, spherically symmetric thin--shell wormholes supported by a generalized Chaplygin gas were theoretically constructed by using the usual cut and paste procedure for a general class of metrics. Such kind of fluid has received great attention in cosmology in the last few years, because it provides a possible explanation for the accelerated expansion of the Universe, and it has also been considered in previous wormhole studies. For a general class of metrics, the equation that determines the possible radii of the throat for static wormholes was obtained and the stability of the static configurations under radial perturbations was analyzed using the standard potential method. The energy density and the pressure at the throat were obtained as functions of the throat radius. Examples of wormholes made from Reissner--Nordstr\"om and Schwarzschild with a cosmological constant metrics were analyzed in detail and the results were compared with those obtained in a previous work \cite{chaply} for the original Chaplygin gas. It was found that for properly chosen values of the parameters, stable solutions are also possible when the gas exponent $\alpha $ is smaller than $1$. The main difference when $0<\alpha <1$ is the presence of an extra unstable solution which appears for some values of the parameters in both charged and anti--de Sitter wormholes. In the charged case it happens when the charge $|Q|$ is slightly smaller than the mass $M$, while in the anti--de Sitter case for large values of $|\Lambda| M^2$.

\section*{Acknowledgments}

This work has been supported by Universidad de Buenos Aires and CONICET.


\begin{thebibliography}{99}

\bibitem{motho} M.S. Morris and K.S. Thorne, Am. J. Phys. 
\textbf{56}, 395 (1988).

\bibitem{visser} M. Visser, \textit{Lorentzian Wormholes} (AIP Press, New
York, 1996).

\bibitem{hovis1} D. Hochberg and M. Visser, Phys. Rev. D \textbf{56}, 4745
(1997).

\bibitem{hovis2} D. Hochberg and M. Visser, Phys. Rev. Lett. \textbf{81}, 746
(1998); D. Hochberg and M. Visser, Phys. Rev D \textbf{58}, 044021 (1998).

\bibitem{viskardad} M. Visser, S. Kar and N. Dadhich, Phys. Rev. Lett. 
\textbf{90}, 201102 (2003).

\bibitem{dil} E. F. Eiroa and C. Simeone, Phys. Rev. D \textbf{71}, 127501 (2005).

\bibitem{lst} O.B. Zaslavskii, Phys. Rev. D \textbf{76}, 044017 (2007).

\bibitem{mvis} M. Visser, Phys. Rev. D \textbf{39}, 3182 (1989); 
M. Visser, Nucl. Phys. \textbf{B328}, 203 (1989).

\bibitem{poisson} E. Poisson and M. Visser, Phys. Rev. D \textbf{52}, 7318
(1995).

\bibitem{barcelo} C. Barcel\'{o} and M. Visser, Nucl. Phys. \textbf{B584}, 415 
(2000). 

\bibitem{ishak} M. Ishak and K. Lake, Phys. Rev. D \textbf{65}, 044011 (2002).

\bibitem{eirom} E.F. Eiroa and G.E. Romero, Gen. Relativ. Gravit. \textbf{36}, 
651 (2004).

\bibitem{lobo} F.S.N. Lobo and P. Crawford, Class. Quantum Grav. \textbf{21}, 
391 (2004). 

\bibitem{lobo2} F.S.N. Lobo and P. Crawford, Class. Quantum Grav. \textbf{22}, 
4869 (2005). 

\bibitem{marc} M. Thibeault, C. Simeone and E. F. Eiroa, Gen. Relativ. Gravit. \textbf{38}, 1593 (2006). 

\bibitem{eir} E.F. Eiroa, Phys. Rev. D \textbf{78}, 024018 (2008).

\bibitem{eisi} E. F. Eiroa and C. Simeone, Phys. Rev. D \textbf{70}, 044008 (2004); C. Bejarano, E. F. Eiroa and C. Simeone, Phys. Rev. D \textbf{75}, 027501 (2007); M.G. Richarte and C. Simeone, Phys. Rev. D \textbf{79}, 127502 (2009).

\bibitem{other}  F. Rahaman, M. Kalam, and S. Chakraborti, Int. J. Mod. Phys. D \textbf{16}, 1669 (2007); E. Gravanis and S. Willison, Phys. Rev. D \textbf{75}, 084025 (2007); M.G. Richarte and C. Simeone, Phys. Rev. D \textbf{76}, 087502 (2007); \textit{ibid.} \textbf{77}, 089903 (E) (2008); C. Garraffo and G. Giribet, E. Gravanis, S. Willison, J. Math. Phys. \textbf{49}, 042502 (2008); C. Garraffo and G. Giribet, Mod. Phys. Lett. A \textbf{23}, 1801 (2008); J.P.S. Lemos and F.S.N. Lobo, Phys. Rev. D \textbf{78}, 044030 (2008); E. F. Eiroa and C. Simeone, Phys. Lett. A \textbf{373}, 1 (2008), \textit{ibid.} \textbf{373}, 2399 (E) (2009); K.A. Bronnikov and A.A. Starobinsky, Mod. Phys. Lett. A \textbf{24}, 1559 (2009). 

\bibitem{acc} A. Riess et al., Astron. J. \textbf{116}, 1009 (1998); S. J. Perlmutter et al., Astroph. J. \textbf{517}, 565 (1999); N. A. Bahcall, J. P. Ostriker,  S. J. Perlmutter and P. J. Steinhardt, Science \textbf{284}, 1481 (1999).

\bibitem{matt} V. Sahni and A. A. Starobinsky, Int. J. Mod. Phys. D \textbf{9}, 373 (2000); P. J. Peebles and B. Ratra, Rev. Mod. Phys. \textbf{75}, 559 (2003); T. Padmanabhan, Phys. Rep. \textbf{380}, 235 (2003).

\bibitem{chap} A. Kamenshchik, U. Moschella and V. Pasquier, Phys. Lett. \textbf{B511}, 265 (2001); N. Bili\'{c}, G.B. Tupper and R.D. Viollier, Phys. Lett. \textbf{B535}, 17 (2002); M.C. Bento, O. Bertolami and A.A. Sen, Phys. Rev. D \textbf{66}, 043507 (2002); for a review see V. Gorini, A. Kamenshchik, U. Moschella and V. Pasquier, gr-qc/0403062.

\bibitem{aero} S. Chaplygin, Sci. Mem. Moscow Univ. Math. Phys. \textbf{21}, 1 (1904); H.-S-Tien, J. Aeron. Sci. \textbf{6}, 399 (1939); T. von Karman,  J. Aeron. Sci. \textbf{8}, 337 (1941).
 
\bibitem{brane} J. M. Bordemann and J. Hoppe, Phys. Lett \textbf{B317}, 315 (1993).

\bibitem{phantom} P. F. Gonz\'{a}lez-D\'{\i}az, Phys. Rev. Lett. \textbf{93}, 071301 (2004); S. V. Sushkov, Phys. Rev. D \textbf{71}, 043520 (2005); V. Faraoni and W. Israel, Phys. Rev. D \textbf{71}, 064017 (2005); F. S. N. Lobo, Phys. Rev. D \textbf{71}, 084011 (2005);  \textit{ibid.} \textbf{71}, 124022 (2005); F. Rahaman, M. Kalam, N. Sarker and K. Gayen, Phys. Lett. \textbf{B633}, 161 (2006); K. A. Bronnikov and J. C. Fabris, Phys. Rev. Lett. \textbf{96}, 251101 (2006); K. A. Bronnikov and A. A. Starobinsky, JETP Lett. \textbf{85}, 1 (2007) [Pis'ma Zh. Eksp. Teor. Fiz. \textbf{85}, 3 (2007)]; A. DeBenedictis, R. Garattini and F.S.N. Lobo, Phys. Rev. D \textbf{78}, 104003 (2008); M. Cataldo, P. Labrana, S. del Campo, J. Crisostomo and P. Salgado, Phys. Rev. D \textbf{78}, 104006 (2008);  M. Cataldo, S. del Campo, P. Minning and P. Salgado, Phys. Rev. D \textbf{79}, 024005 (2009); M. Jamil, P.K.F. Kuhfittig, F. Rahaman and S.A. Rakib, arXiv:0906.2142 [gr-qc]; J. A. Gonzalez, F. S. Guzman, N. Montelongo-Garcia, and T. Zannias, Phys. Rev. D \textbf{79}, 064027 (2009). 

\bibitem{lobo73} F. S. N. Lobo, Phys. Rev. D \textbf{73}, 064028 (2006).

\bibitem{chaply} E.F. Eiroa and C. Simeone, Phys. Rev. D \textbf{76}, 024021 (2007).

\bibitem{chapnew} F. Rahaman, M. Kalam and K. A. Rahman, Mod. Phys. Lett. A \textbf{23}, 1199 (2008); V. Gorini, U. Moschella, A.Yu. Kamenshchik, V. Pasquier, and A.A. Starobinsky, Phys. Rev. D \textbf{78}, 064064 (2008); S. Chakraborty and T. Bandyopadhyay, Int. J. Mod. Phys.D \textbf{18}, 463 (2009); M. Jamil, M. U. Farooq and M. A. Rashid, Eur. Phys. J. C \textbf{59}, 907 (2009); P.K.F. Kuhfittig, Gen. Relativ. Grav. \textbf{41}, 1485 (2009). 

\bibitem{daris} N. Sen, Ann. Phys. (Leipzig) \textbf{73}, 365 (1924); K.
Lanczos, \textit{ibid.} \textbf{74}, 518 (1924); G. Darmois, M\'{e}morial des
Sciences Math\'{e}matiques, Fascicule XXV, Chap. V (Gauthier-Villars, Paris, 1927)
; W. Israel, Nuovo Cimento \textbf{44B}, 1 (1966); \textbf{48B}, 463(E) (1967); for a review see P. Musgrave and K. Lake, Class. Quantum Grav. \textbf{13}, 1885
(1996).

\end{thebibliography}
\end{document}